\documentclass[preprint,prd,aps,showpacs,showkeys,nofootinbib]{revtex4}
\usepackage{graphicx}
\usepackage{dcolumn}
\usepackage{bm}
\usepackage{color}
\usepackage[dvipsnames]{xcolor}
\usepackage{amssymb}
\usepackage{epstopdf}

\definecolor{black-blue}{RGB}{77,116,175}
\definecolor{black-yellow}{RGB}{231,162,33}
\definecolor{black-green}{RGB}{144,180,58}
\definecolor{black-red}{RGB}{246,95,50}

\begin{document}

\title{Calculation for Electric Dipole Moments of Lepton and Neutron in the N-B-LSSM via the Mass Insertion Approximation}
\author{Shuang Di$^{1,2,3}$, Wei-Hang Zhang$^{1,2,3}$, Rong-Zhi Sun$^{1,2,3}$, Xing-Xing Dong$^{1,2,3,4}$\footnote{dongxx@hbu.edu.cn}, Guo-Zhu Ning$^{1,2,3}$\footnote{ninggz@hbu.edu.cn}, Shu-Min Zhao$^{1,2,3}$\footnote{zhaosm@hbu.edu.cn}}

\affiliation{$^1$ Department of Physics, Hebei University, Baoding 071002, China}
\affiliation{$^2$ Hebei Key Laboratory of High-precision Computation and Application of Quantum Field Theory, Baoding, 071002, China}
\affiliation{$^3$ Hebei Research Center of the Basic Discipline for Computational Physics, Baoding, 071002, China}
\affiliation{$^4$ Departamento de Fisica and CFTP, Instituto Superior T$\acute{e}$cnico, Universidade de Lisboa,
Av.Rovisco Pais 1,1049-001 Lisboa, Portugal}
\date{\today}

\begin{abstract}
In the N-B-LSSM, we calculate the electric dipole moments (EDMs) of lepton and neutron at the one loop level via the Mass Insertion Approximation (MIA). In the Standard Model (SM), charge parity (CP) violation originates only from the single phase of the Cabibbo-Kobayashi-Maskawa (CKM) matrix, and the predicted EDMs of lepton and neutron are far below the current experimental upper limits. Thus, EDMs serve as sensitive probes for exploring CP-violating phases in new physics.

The N-B-LSSM extends the Minimal Supersymmetric Standard Model (MSSM) by introducing right-handed neutrino superfields and additional singlet Higgs superfields, which enriches the particle spectrum and the sources of CP violation. We derive the one loop analytical expressions for lepton and quark EDMs, and reveal their dependence on model parameters such as $g_{YB}$, $\theta_{\mu_H}$, $\theta_{1'}$, $\theta_{BB'}$ and $\tan\beta$. Numerical analyses demonstrate that EDM measurements strongly constrain the CP-violating parameter space of the N-B-LSSM and reveal the sensitivity of lepton and neutron EDMs to the extended gauge interactions and supersymmetric parameters.
This study provides a systematic theoretical tool and numerical reference for exploring CP violation and new physics under the N-B-LSSM.
\end{abstract}


\maketitle

\section{Introduction}
The SM of particle physics is highly successful. In the SM, the only source of CP violation comes from a single phase in the CKM matrix\cite{A1,A2,C21,C22,C23}. This source predicts extremely small EDMs for fermions. The predicted values are far below current experimental limitations. This promotes new physics beyond the SM (BSM). BSM models can add new CP-violating phases, and those
 phases can enhance the EDMs of the lepton and neutron\cite{A3}.

In recent years, the improvements in experimental techniques have significantly enhanced constraints on the EDMs of leptons and neutron. The current limits on the EDMs of leptons and neutron are as follows\cite{A4,A5,A6,A7,A8}
\begin{eqnarray}
&&\hspace{-2cm}\mid{d_e}\mid < 4.1\times10^{-30}~\rm{e.cm} ,~CL = 90\%
\nonumber\\&&\hspace{-2cm}\mid{d_\mu}\mid < 1.8\times10^{-19}~\rm{e.cm} ,~CL = 95\%
\nonumber\\&&\hspace{-2cm}\mid{d_\tau}\mid <1.03 \times10^{-17}~\rm{e.cm} ,~CL = 95\%
\nonumber\\&&\hspace{-2cm}\mid{d_n}\mid < 1.8\times10^{-26}~\rm{e.cm} ,~CL = 90\%
\end{eqnarray}
These strict experimental results set strong limits on the new CP-violation mechanisms in BSM models and also make it important to calculate the EDMs in the new models.

Supersymmetry (SUSY) is a well-known extension of the SM. SUSY naturally introduces new CP-violating phases. Examples include the phase $\theta_3$ of the gluino mass term $m_g$, and the phase $\theta_{\mu_H}$ of the Higgsino mass term $\mu_H$. These phases can support electroweak baryogenesis (EWB)\cite{A9,A10,A11,C111,C112,C113}. At the same time, they can strongly enhance lepton and neutron EDMs\cite{A12}. Therefore, achieving sufficient CP violation while suppressing EDM contributions remains a central issue in SUSY model building\cite{A13}.

This work considers extensions of the MSSM equipped with a local $U(1)_{B-L}$ gauge symmetry, which introduces a much broader symmetry structure. Our central object of study is the N-B-LSSM, a next-to-minimal supersymmetric theory with an intrinsic $B-L$ gauge sector, whose gauge group reads $SU(3)_C \times SU(2)_L \times U(1)_Y \times U(1)_{B-L}.$
This model introduces right-handed neutrino superfields and additional singlet Higgs superfields on the basis of the Minimal Supersymmetric Standard Model (MSSM). On the one hand, light neutrino masses can be explained by the seesaw mechanism\cite{D14}. On the other hand, the vacuum expectation value (VEV) of the singlet field can generate an effective $\mu$ term, thereby alleviating the $\mu$ problem\cite{H15}. Importantly, gauge kinetic mixing between the two $U(1)$ factors is a characteristic feature of this class of models\cite{A16}. It modifies the mixing structure and interactions of gauge bosons and neutral gauginos, and thus can have an impact on the main EDM contributions\cite{A17}.

In the N-B-LSSM, new physics contributions to lepton and quark EDMs are mainly induced by supersymmetric loop diagrams. Lepton EDMs receive contributions from neutralino $-$ slepton and chargino $-$ sneutrino loops. Quark EDMs and CEDMs are dominated by gluino $-$ squark loops as well as neutralino $-$ squark loops\cite{A5,A12}. Extended degrees of freedom include additional neutralinos and singlet Higgs states. New CP-violating phases link to soft-breaking parameters and effective $\mu$ couplings parameters. These two components significantly modify the magnitude of EDMs\cite{Y18}. These additional CP-violating sources enrich the EDM phenomenology of the N-B-LSSM and allow EDM observables to probe the characteristic features of the extended gauge and neutralino sectors. Meanwhile, current EDM measurements provide stringent constraints on the CP-violating parameter space of the model \cite{A3,A5,A6,Basso2009}.

This paper aims to compute lepton EDMs and the neutron EDM (induced by quark EDM and CEDM) in the N-B-LSSM using the MIA. The MIA provides an intuitive understanding of parameter dependence, which helps to identify the main sources affecting the EDM predictions. In particular, we focus on some CP-violating phases, and systematically analyze the impact of parameters including $g_B$, $g_{YB}$, and the additional soft mass terms $m_{BB'}$ and $m_{B'}$ on the EDMs. Through numerical scans, we show that there exists a viable parameter space in which the EDMs of lepton and neutron satisfy the current experimental upper bounds, thereby providing effective constraints on the model parameters.

The structure of this paper is as follows: In Section 2, we briefly review the N-B-LSSM. Section 3 details the calculation of EDMs for lepton and neutron using the mass insertion approach. In Section 4, we present numerical results and discuss their implications for the EDMs of lepton and neutron, along with the relevant physical explanations. Finally, conclusions are provided in Section 5.
\section{The $N-B-LSSM$}
The local gauge group of N-B-LSSM is \cite{MX1}
\begin{equation}
SU(3)_C \times SU(2)_L \times U(1)_Y \times U(1)_{B-L}.
\end{equation}
Compared with the MSSM, the model introduces additional superfields,
including three singlet Higgs superfields $\hat{\chi}_1$, $\hat{\chi}_2$, $\hat{S}$,
together with the corresponding fermionic partners (Higgsinos),
and an additional $U(1)_{B-L}$ gaugino $\lambda_{\widetilde{B'}}$.
After symmetry breaking, the singlet field $S$ develops a nonzero VEV.
Consequently, the superpotential term $\lambda \hat S \hat H_u \hat H_d$ generates an effective $\mu$ parameter,
\begin{equation}
\mu_{\rm eff}=\frac{\lambda v_S}{\sqrt{2}},
\end{equation}
which alleviates the $\mu$-problem of the MSSM.
Throughout this work we assume $R$-parity conservation, which is automatically preserved in the N-B-LSSM due to the gauged $B-L$ symmetry.
With the singlet sector, the neutral CP-even scalar fields of
$H_u$, $H_d$, $\chi_1$, $\chi_2$ and $S$ mix into a $5\times5$ mass squared matrix.
The most significant impact for EDM is that the additional singlet/gauge sectors alter the neutralino (and Higgs) mass spectra and the associated couplings, thereby introducing new CP-sensitive contributions to the one loop EDM amplitudes.

The superpotential of the N-B-LSSM is
\begin{eqnarray}
W&=&-Y_d\hat{d}\hat{q}\hat{H}_d-Y_e\hat{e}\hat{l}\hat{H}_d-\lambda_2\hat{S}\hat{\chi}_1\hat{\chi}_2+\lambda\hat{S}\hat{H}_u\hat{H}_d
\nonumber\\&&\hspace{-1.5cm}~~~~~~~+\frac{\kappa}{3}\hat{S}\hat{S}\hat{S}+Y_u\hat{u}\hat{q}\hat{H}_u+Y_{\chi}\hat{\nu}\hat{\chi}_1\hat{\nu}+Y_\nu\hat{\nu}\hat{l}\hat{H}_u.
\end{eqnarray}
Here $Y_{u,d,e,\nu,\chi}$ denote Yukawa coupling matrices, while $\lambda$, $\lambda_2$, and $\kappa$ are dimensionless couplings.
The fields $\hat{\chi}_1$, $\hat{\chi}_2$, and $\hat{S}$ are singlet Higgs superfields.
Note that a term of the form $Y'_\nu\hat{\nu}\hat{l}\hat{S}$ is forbidden by gauge invariance,
since the sum of the corresponding $U(1)_Y$ charges of $\hat{\nu}$, $\hat{l}$ and $\hat{S}$ is nonzero.
In the following EDM analysis, the gaugino/soft terms
may carry complex phases after field redefinitions, providing potential sources of CP violation.

The VEVs of the Higgs fields $H_u$, $H_d$, $\chi_1$, $\chi_2$, and $S$
are denoted by $v_u$, $v_d$, $v_{\eta}$, $v_{\bar{\eta}}$, and $v_S$, respectively.
Two angles are defined as $\tan\beta=v_u/v_d$ and $\tan\beta_\eta=v_{\bar{\eta}}/v_{\eta}$.
The explicit expressions of the neutral parts for the two Higgs doublets and three Higgs singlets around their VEVs are
\begin{eqnarray}
H^0_d&=&\frac{1}{\sqrt{2}}\phi_{d}+\frac{1}{\sqrt{2}}v_{d}+i\frac{1}{\sqrt{2}}\sigma_d,
\nonumber\\
H^0_u&=&\frac{1}{\sqrt{2}}\phi_{u}+\frac{1}{\sqrt{2}}v_{u}+i\frac{1}{\sqrt{2}}\sigma_u,
\nonumber\\
\chi_1&=&\frac{1}{\sqrt{2}}\phi_{1}+\frac{1}{\sqrt{2}}v_{\eta}+i\frac{1}{\sqrt{2}}\sigma_1,
\nonumber\\
\chi_2&=&\frac{1}{\sqrt{2}}\phi_{2}+\frac{1}{\sqrt{2}}v_{\bar{\eta}}+i\frac{1}{\sqrt{2}}\sigma_2,
\nonumber\\
S&=&\frac{1}{\sqrt{2}}\phi_S+\frac{1}{\sqrt{2}}v_S+i\frac{1}{\sqrt{2}}\sigma_S.
\end{eqnarray}
In the EDM calculation, the above VEVs affect the neutralino/Higgs mixings and thus enter the
couplings relevant for the one loop neutralino--slepton and chargino--sneutrino contributions.

The soft SUSY-breaking terms of the N-B-LSSM are
\begin{eqnarray}
\mathcal{L}_{soft}&=&\mathcal{L}_{soft}^{MSSM}-\frac{T_\kappa}{3}S^3+\epsilon_{ij}T_{\lambda}SH_d^iH_u^j+T_{2}S\chi_1\chi_2\nonumber\\
&&-T_{\chi,ik}\chi_1\tilde{\nu}_{R,i}^{*}\tilde{\nu}_{R,k}^{*}
+\epsilon_{ij}T_{\nu,ij}H_u^i\tilde{\nu}_{R,i}^{*}\tilde{e}_{L,j}
-m_{\eta}^2|\chi_1|^2-m_{\bar{\eta}}^2|\chi_2|^2\nonumber\\
&&-m_S^2|S|^2-m_{\nu,ij}^2\tilde{\nu}_{R,i}^{*}\tilde{\nu}_{R,j}
-\frac{1}{2}\Big(2M_{BB^\prime}\tilde{B}\lambda_{\tilde{B^\prime}}+\delta_{ij} M_{BL}\lambda_{\tilde{B^\prime}}^2\Big)+h.c.~.
\end{eqnarray}

Here $\mathcal{L}_{\text{soft}}^{\text{MSSM}}$ denotes the soft-breaking terms of the MSSM, with $T_\kappa$, $T_\lambda$, $T_2$, $T_\chi$ and $T_\nu$ being the trilinear soft-breaking parameters. In our EDM analysis, the dominant CP-violating phases stem from complex soft-breaking parameters, specifically the gaugino mass parameters $M_1$, $M_2$, $M_{BL}$ and the mixing mass $M_{BB'}$, and complex trilinear terms. $\lambda_{\tilde{B}'}$ denotes the gaugino field associated with the additional $U(1)_{B-L}$ gauge symmetry. This field extends the neutralino sector and participates in the mixing induced by the $U(1)_Y-U(1)_{B-L}$ gauge kinetic mixing. These phases induce CP-odd couplings and chirality flips in loop amplitudes, and the relevant chirality-changing insertions are introduced within the framework of the MIA.

\begin{table}[h]
\centering
\caption{The superfields in N-B-LSSM.}
\begin{tabular}{|c|c|c|c|c|}
\hline
Superfields & $U(1)_Y$ & $SU(2)_L$ & $SU(3)_C$ & $U(1)_{B-L}$ \\
\hline
$\hat{q}$        & $1/6$   & $2$ & $3$        & $1/6$   \\
\hline
$\hat{l}$        & $-1/2$  & $2$ & $1$        & $-1/2$  \\
\hline
$\hat{H}_d$      & $-1/2$  & $2$ & $1$        & $0$     \\
\hline
$\hat{H}_u$      & $1/2$   & $2$ & $1$        & $0$     \\
\hline
$\hat{d}$        & $1/3$   & $1$ & $\bar{3}$  & $-1/6$  \\
\hline
$\hat{u}$        & $-2/3$  & $1$ & $\bar{3}$  & $-1/6$  \\
\hline
$\hat{e}$        & $1$     & $1$ & $1$        & $1/2$   \\
\hline
$\hat{\nu}$      & $0$     & $1$ & $1$        & $1/2$   \\
\hline
$\hat{\chi}_1$   & $0$     & $1$ & $1$        & $-1$    \\
\hline
$\hat{\chi}_2$   & $0$     & $1$ & $1$        & $1$     \\
\hline
$\hat{S}$        & $0$     & $1$ & $1$        & $0$     \\
\hline
\end{tabular}
\label{table1}
\end{table}

The particle content and charge assignments are summarized in Table~\ref{table1}.
In the chiral superfield notation,
$\hat H_u = (\hat H_u^+,\hat H_u^0)^T$ and $\hat H_d = (\hat H_d^0,\hat H_d^-)^T$
denote the MSSM-like Higgs doublet superfields.
The superfields $\hat q$ and $\hat l$ represent the left-handed quark and lepton doublets,
while $\hat u$, $\hat d$, $\hat e$, and $\hat\nu$ correspond to the right-handed up-type quark, down-type quark,
charged lepton, and neutrino singlet superfields, respectively.

The gauge groups $U(1)_Y$ and $U(1)_{B-L}$ generally exhibit gauge kinetic mixing,
which can be induced through RGEs even if it is set to zero at $M_{GUT}$. A basis transformation can be performed by an orthogonal rotation matrix $R$ ($R^TR=1$)\cite{A17,MX3,MX4,MX5}.
The covariant derivative can be expressed as
\begin{eqnarray}
D_\mu&=&\partial_\mu-i\left(\begin{array}{cc}Y,&B-L\end{array}\right)
\left(\begin{array}{cc}g_{Y},&g{'}_{{YB}}\\g{'}_{{BY}},&g{'}_{{B-L}}\end{array}\right)
\left(\begin{array}{c}B_{\mu}^{\prime Y} \\ B_{\mu}^{\prime BL}\end{array}\right)\;,
\label{gauge1}
\end{eqnarray}
where $Y$ and $B-L$ denote the hypercharge and the $B-L$ charge, respectively.
One may choose a basis such that
\begin{eqnarray}
\left(\begin{array}{cc}g_{Y},&g{'}_{{YB}}\\g{'}_{{BY}},&g{'}_{{B-L}}\end{array}\right)
R^T&=&\left(\begin{array}{cc}g_{1},&g_{{YB}}\\0,&g_{{B}}\end{array}\right)\;,
\label{gauge3}
\end{eqnarray}
and consequently the $U(1)$ gauge fields are redefined as
\begin{eqnarray}
R\left(\begin{array}{c}B_{\mu}^{\prime Y} \\ B_{\mu}^{\prime BL}\end{array}\right)
&=&\left(\begin{array}{c}B_{\mu}^{Y} \\ B_{\mu}^{BL}\end{array}\right)\;.
\label{gauge4}
\end{eqnarray}

In the neutralino sector, the mass matrix in the basis
$(\tilde{B}, \tilde{W}^0, \tilde{H}_d^0, \tilde{H}_u^0,
\lambda_{\tilde{B'}}, \tilde{\chi}_1, \tilde{\chi}_2, \tilde{S})$
is given by
{\small\begin{eqnarray}
m_{\chi^0} &=& \left(
\begin{array}{cccccccc}
M_1 &0 &-\frac{1}{2}g_1 v_d &\frac{1}{2} g_1 v_u &{M}_{B B'} & 0  & 0  &0\\
0 &M_2 &\frac{1}{2} g_2 v_d  &-\frac{1}{2} g_2 v_u  &0 &0 &0 &0\\
-\frac{1}{2}g_1 v_d &\frac{1}{2} g_2 v_d  &0
&- \frac{1}{\sqrt{2}} {\lambda} v_S&-\frac{1}{2} g_{YB} v_d &0 &0 & - \frac{1}{\sqrt{2}} {\lambda} v_u\\
\frac{1}{2}g_1 v_u &-\frac{1}{2} g_2 v_u  &- \frac{1}{\sqrt{2}} {\lambda} v_S &0 &\frac{1}{2} g_{YB} v_u  &0 &0 &- \frac{1}{\sqrt{2}} {\lambda} v_d\\
{M}_{B B'} &0 &-\frac{1}{2} g_{YB} v_{d}  &\frac{1}{2} g_{YB} v_{u} &{M}_{BL} &- g_{B} v_{\eta}  &g_{B} v_{\bar{\eta}}  &0\\
0  &0 &0 &0 &- g_{B} v_{\eta}  &0 &-\frac{1}{\sqrt{2}} {\lambda}_{2} v_S  &-\frac{1}{\sqrt{2}} {\lambda}_{2} v_{\bar{\eta}} \\
0 &0 &0 &0 &g_{B} v_{\bar{\eta}}  &-\frac{1}{\sqrt{2}} {\lambda}_{2} v_S  &0 &-\frac{1}{\sqrt{2}} {\lambda}_{2} v_{\eta} \\
0 &0 & - \frac{1}{\sqrt{2}} {\lambda} v_u &- \frac{1}{\sqrt{2}}{\lambda} v_d &0 &-\frac{1}{\sqrt{2}} {\lambda}_{2} v_{\bar{\eta}}
 &-\frac{1}{\sqrt{2}} {\lambda}_{2} v_{\eta}  &\sqrt{2}\kappa v_S
\end{array}
\right),
\label{neutralino}
\end{eqnarray}}
This matrix is diagonalized by a unitary matrix $N$,
\begin{equation}
N^{*}m_{\chi^0} N^{\dagger} = m^{\rm diag}_{\chi^0}.
\end{equation}
\section{ANALYTICAL FORMULA}
\subsection{The Lepton EDM}
Using the on-shell condition for the external leptons, the lepton EDM can be obtained through the effective Lagrangian as follows\cite{EL1,EL2,EL3}
 \begin{eqnarray}
&&{\cal L}_{{EDM}}=-{i\over2}d_{l}\overline{l}\sigma^{\mu\nu}\gamma_5
lF_{{\mu\nu}}
\label{eq1}.
\end{eqnarray}
where \(F_{\mu\nu}\) is the electromagnetic field strength tensor, and \(l\) denotes the lepton field. This effective Lagrangian explicitly violates CP symmetry and therefore cannot appear at tree level in the fundamental interactions. However, in a CP-violating electroweak theory, it can be generated at least at the one-loop level.

The Feynman amplitude corresponding to the process $ l_j \to l_i + \gamma$ is dominated by a set of CP-violating dimension-6 operators, whose contributions are much larger than those of higher-dimensional operators, which are suppressed by a factor of $m^2_l / M^2_{\text{SUSY}} \sim (10^{-7}, 10^{-8})$. After imposing the on-shell conditions for the external leptons, only the specific operator combinations $ O^{L,R}_{2,3,6}$ give direct contributions to the lepton EDM, with their magnitudes determined by the corresponding Wilson coefficients $C^{L,R}_{2,3,6}$. These operators are constructed from the covariant derivative $D_\mu = \partial_\mu + i e A_\mu $ and the chiral projection operators $ P_{L,R} = (1 \mp \gamma_5)/2 $. The specific forms of those dimension 6 operators are\cite{6D}
\begin{eqnarray}
&&\mathcal{O}_1^{L,R}=\frac{1}{(4\pi)^2}\bar{l}(i\mathcal{D}\!\!\!\slash)^3P_{L,R}l,~~~~~~~~~~~~~~
\mathcal{O}_2^{L,R}=\frac{eQ_f}{(4\pi)^2}\overline{(i\mathcal{D}_{\mu}l)}\gamma^{\mu}
F\cdot\sigma P_{L,R}l,
\nonumber\\
&&\mathcal{O}_3^{L,R}=\frac{eQ_f}{(4\pi)^2}\bar{l}F\cdot\sigma\gamma^{\mu}
P_{L,R} (i\mathcal{D}_{\mu}l),~~~~\mathcal{O}_4^{L,R}=\frac{eQ_f}{(4\pi)^2}\bar{l}(\partial^{\mu}F_{\mu\nu})\gamma^{\nu}
P_{L,R}l,\nonumber\\&&
\mathcal{O}_5^{L,R}=\frac{m_l}{(4\pi)^2}\bar{l}(i\mathcal{D}\!\!\!\slash)^2P_{L,R}l,
~~~~~~~~~~~~~~\mathcal{O}_6^{L,R}=\frac{eQ_fm_l}{(4\pi)^2}\bar{l}F\cdot\sigma
P_{L,R}l.\label{operators}
\end{eqnarray}

\begin{figure}[ht]
\setlength{\unitlength}{5.0mm}
\centering
\includegraphics[width=4.8in]{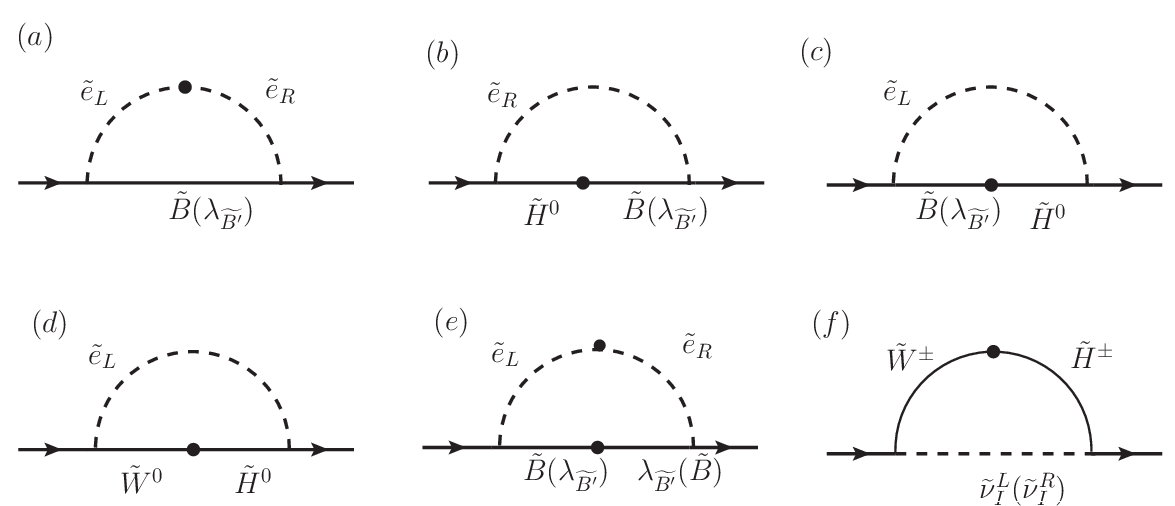}
\caption{Feynman diagrams for generating lepton EDM based on MIA. The external photons are connected to the charged internal lines in all possible ways.}\label{T1}
\end{figure}

In the N-B-LSSM, the one-loop corrections to the lepton electric dipole moment primarily originate from loop diagrams involving sleptons and gauginos (such as the neutralino $\chi^0$ and the chargino $\chi^\pm$) with new physics parameters. Moreover, due to the introduction of right-handed neutrino superfields, the mass spectrum and mixing structure of the sneutrinos become more complicated and may provide additional sources of CP violation. The EDM contributions are generally not proportional to a single mass ratio but are instead constrained by mixing matrix elements. Their magnitudes are highly sensitive to new CP-violating phases and parameters such as $\tan\beta$ in the model. To clearly reveal this parameter dependence, it is particularly convenient to employ the MIA\cite{MIA}, which expresses the EDM contributions analytically as functions of the model parameters, thereby highlighting the origin of CP violation. Fig.\ref{T1} shows the Feynman diagrams for the lepton EDM induced from MIA.

The MIA provides a powerful framework for
analyzing supersymmetric CP-violating observables. The EDM contributions
from neutralino-slepton, chargino-sneutrino, and gluino--squark loop
diagrams have been extensively studied within the MSSM framework \cite{MSSM1,MSSM2,MSSM3}. Compared with the MSSM, N-B-LSSM has new CP-violating sources including the additional $U(1)_{B-L}$
gaugino mass ($M_{BL}$), the mass mixing term($M_{BB'}$) of $U(1)_{Y}$ gaugino and $U(1)_{B-L}$ gaugino,
the mass (mixing) terms of the supper partners
for the Higgs singlets $\chi_1$ $\chi_2$ and $S$.
Furthermore, the new gauge coupling
parameters $g_B$ and $g_{YB}$ can affect the EDM corrections obviously.
In the following, within the N-B-LSSM and using the MIA, we show the explicit analytical forms
of the one-loop contributions to the lepton EDM.

a. The one-loop contributions from $\widetilde{B}$-$\tilde{e}_L$-$\tilde{e}_R$ and $\lambda_{\tilde{B'}}$-$\tilde{e}_L$-$\tilde{e}_R$.
\begin{eqnarray}
&&d_l^{(a)}(\widetilde{e}_L,\widetilde{e}_R,\widetilde{B})=\frac{eg_1^2}{M_{SUSY}}\sqrt{x_1x_lx_{\mu_H}}
e^{i\theta_1}e^{i\theta_{\mu_H}}\tan\beta I_1(x_1,x_{\widetilde{e}_L},x_{\widetilde{e}_R}).
\\&&
d_l^{(a)}(\widetilde{e}_L,\widetilde{e}_R,\lambda_{\widetilde{B'}})=\frac{e(g_{YB}+g_B)(2g_{YB}+g_B)}{2M_{SUSY}}\sqrt{x_{\lambda_{\widetilde{B'}}}x_lx_{\mu_H}}
\nonumber\\&&\times e^{i\theta_1'}e^{i\theta_{\mu_H}}\tan\beta I_1(x_{\lambda_{\widetilde{B'}}},x_{\widetilde{e}_L},x_{\widetilde{e}_R}).
\end{eqnarray}
Here $x_i=\frac{|m_i|^2}{M^2_{SUSY}}$ with $M_{SUSY}$ denoting the supersymmetric mass scale.

The function $I_1(x,y,z)$ is defined as
\begin{eqnarray}
I_1(x,y,z)= \frac{1}{32\pi^2}\Big\{\frac{-3 x^2 + y z + x (y+z)}{(x-y)^2 (x-z)^2} + \frac{2 x [x^3 - 3 x y z + y z (y+z)] \log x}{(x-y)^3 (x-z)^3}
\nonumber\\&&\hspace{-12cm}-\frac{2 x y \log y}{(x-y)^3 (y-z)} - \frac{2 x z \log z}{(x-z)^3 (z-y)}\Big\}.
\end{eqnarray}

b. The one-loop contributions from $\widetilde{B}$-$\widetilde{H}^0$-$\tilde{e}_R$ and $\lambda_{\tilde{B'}}$-$\widetilde{H}^0$-$\tilde{e}_R$.
\begin{eqnarray}
&&d_l^{(b)}(\widetilde{e}_R,\widetilde{H}^0,\widetilde{B})=\frac{eg_1^2}{M_{SUSY}}\sqrt{x_1x_lx_{\mu_H}}e^{i\theta_1}e^{i\theta_{\mu_H}}\tan\beta I_2(x_1,x_{\mu_H},x_{\widetilde{e}_R}).
\\&&
d_l^{(b)}(\widetilde{e}_R,\widetilde{H^0},\lambda_{\widetilde{B'}})=\frac{eg_1(2g_{YB}+g_B)}{2M_{SUSY}}\sqrt{x_{\lambda_{\widetilde{B'}}}x_lx_{\mu_H}}e^{i\theta_1'}e^{i\theta_{\mu_H}}\tan\beta
I_2(x_{\lambda_{\widetilde{B'}}},x_{\mu_H},x_{\widetilde{e}_R}).
\end{eqnarray}

c. The one-loop contributions from $\widetilde{B}$-$\widetilde{H}^0$-$\tilde{e}_L$ and $\lambda_{\tilde{B'}}$-$\widetilde{H}^0$-$\tilde{e}_L$.
\begin{eqnarray}
&&d_l^{(c)}(\widetilde{e}_L,\widetilde{H^0},\widetilde{B})=-\frac{eg_1^2}{2M_{SUSY}}\sqrt{x_1x_lx_{\mu_H}}e^{i\theta_1}e^{i\theta_{\mu_H}}\tan\beta
I_2(x_1,x_{\mu_H},x_{\widetilde{e}_L}).
\\&&
d_l^{(c)}(\widetilde{e}_L,\widetilde{H}^0,\lambda_{\widetilde{B'}})=-\frac{eg_1(g_{YB}+g_B)}{2M_{SUSY}}\sqrt{x_{\lambda_{\widetilde{B'}}}x_lx_{\mu_H}}e^{i\theta_1'}e^{i\theta_{\mu_H}}\tan\beta
I_2(x_{\lambda_{\widetilde{B'}}},x_{\mu_H},x_{\widetilde{e}_L}).
\end{eqnarray}

d. The one-loop contributions from $\widetilde{W}^0$-$\widetilde{H}^0$-$\tilde{e}_L$.
\begin{eqnarray}
d_l^{(d)}(\widetilde{e}_L,\widetilde{H}^0,\widetilde{W}^0)=\frac{eg_2^2}{2M_{SUSY}}\sqrt{x_2x_lx_{\mu_H}}e^{i\theta_2}e^{i\theta_{\mu_H}}\tan\beta
I_2(x_2,x_{\mu_H},x_{\widetilde{e}_L}).
\end{eqnarray}

The function $I_2(x,y,z)$ is defined as
\begin{eqnarray}
I_2(x,y,z)=\frac{1}{32\pi^2}\Big\{\frac{(y-3z)z + x(y+z)}{(x-z)^2 (y-z)^2} + \frac{2 x z \log x}{(x-y)(x-z)^3}
\nonumber\\&&\hspace{-9cm}+ \frac{2 y z \log y}{(-x+y)(y-z)^3} - \frac{2 z [x y (x+y) - 3 x y z + z^3] \log z}{(x-z)^3 (-y+z)^3}\Big\}.
\end{eqnarray}

e. The one-loop contributions from $\lambda_{\tilde{B'}}$-$\widetilde{B}$-$\tilde{e}_L$-$\tilde{e}_R$ and
$\widetilde{B}$-$\lambda_{\tilde{B'}}$-$\tilde{e}_L$-$\tilde{e}_R$
\begin{eqnarray}
d_l^{(e)}(\widetilde{e}_L,\widetilde{e}_R,\lambda_{\widetilde{B'}},\widetilde{B})=
-\frac{eg_1(g_{YB}+g_B)}{2M_{SUSY}}\sqrt{x_{BB'}x_{\mu_H}}e^{i\theta_{BB'}}e^{i\theta_{\mu_H}}\tan\beta
\nonumber\\&&\hspace{-10cm}\times[\sqrt{x_1x_{\lambda_{\widetilde{B'}}}}
f(x_{BB'},x_{1},x_{\widetilde{e}_L},x_{\widetilde{e}_R})
-g(x_{BB'},x_{1},x_{\widetilde{e}_L},x_{\widetilde{e}_R})],
\\
d_l^{(e)}(\widetilde{e}_L,\widetilde{e}_R,\widetilde{B},\lambda_{\widetilde{B'}})=
-\frac{eg_1(2g_{YB}+g_B)}{4M_{SUSY}}\sqrt{x_{BB'}x_{\mu_H}}e^{i\theta_{BB'}}e^{i\theta_{\mu_H}}\tan\beta
\nonumber\\&&\hspace{-10cm}\times[\sqrt{x_1x_{\lambda_{\widetilde{B'}}}}
f(x_{BB'},x_{1},x_{\widetilde{e}_L},x_{\widetilde{e}_R})
-g(x_{BB'},x_{1},x_{\widetilde{e}_L},x_{\widetilde{e}_R})].
\end{eqnarray}

The functions $f(x,y,z,t)$ and $g(x,y,z,t)$ are defined as
\begin{eqnarray}
f(x,y,z,t) = \frac{1}{16\pi^2} \Big\{
\frac{t [t^3 + x y (x+y-3t)] \log t}{(t-x)^3 (t-y)^3 (t-z)}- \frac{x [x^3 +t z (t+z-3x)] \log x}{(t-x)^3 (x-y) (x-z)^3}
\nonumber\\&&\hspace{-12cm}+ \frac{y [ y^3 +t z (t+z-3y)] \log y}{(t-y)^3 (x-y) (y-z)^3}- \frac{z [z^3+ x y (x+y-3z)] \log z}{(t-z) (z-x)^3 (z-y)^3}
\nonumber\\&&\hspace{-12cm}+ \frac{1}{2(x-y)}
[\frac{t x + t z + x z - 3x^2}{(t-x)^2 (x-z)^2}
- \frac{t y + t z + y z - 3y^2}{(t-y)^2 (y-z)^2}] \Big\};
\end{eqnarray}
\begin{eqnarray}
g(x,y,z,t)=\frac{1}{16\pi^2} \{\frac{x\,(3x^2 -t x - t z - x z -)}{2(t-x)^2 (x-y)(x-z)^2} -\frac{y [t (y+z)+y (z-3 y)]}{2(t-y)^2 (y-x)(y-z)^2}
   \nonumber\\&&\hspace{-12cm}-\frac{t[t^3 (x+y)-3 t^2 x y+x^2y^2] \log t}{(t-x)^3 (t-y)^3 (t-z)}+\frac{z
   [x^2 y^2+x z^2 (z-3 y)+y z^3] \log z}{(t-z)(z-x)^3
   (z-y)^3}
   \nonumber\\&&\hspace{-12cm}+\frac{x^2 [x^3+t z(t+z-3x)]\log x}{(t-x)^3 (x-y)(x-z)^3}-\frac{y^2[y^3+tz(t+z-3y)] \log y}{(t-y)^3 (x-y)(y-z)^3}\}.
\end{eqnarray}

f. The one-loop contributions from $\widetilde{W}^\pm$-$\widetilde{H}^\pm$-$\widetilde{\nu}_L^I$ and $\widetilde{W}^\pm$-$\widetilde{H}^\pm$-$\widetilde{\nu}_L^R$.
\begin{eqnarray}
&&d_l^{(f)}(\widetilde{W}^\pm,\widetilde{H}^\pm,\widetilde{\nu}_L^I)=
-\frac{eg_2^2}{2M_{SUSY}}\sqrt{x_2x_lx_{\mu_H}}e^{i\theta_2}e^{i\theta_{\mu_H}}\tan\beta
I_3(x_2,x_{\mu_H},x_{\nu_L^I}).
\\&&
d_l^{(f)}(\widetilde{W}^\pm,\widetilde{H}^\pm,\widetilde{\nu}^R_L)=
-\frac{eg_2^2}{2M_{SUSY}}\sqrt{x_2x_lx_{\mu_H}}e^{i\theta_2}e^{i\theta_{\mu_H}}\tan\beta
I_3(x_2,x_{\mu_H},x_{\nu^R_L}).
\end{eqnarray}

The function $I_3(x,y,z)$ is defined as
\begin{eqnarray}
&&I_3(x,y,z)=\frac{1}{32\pi^2}\Big\{\frac{2 z^2 [x^2+x (y-3 z)+y^2-3 y z+3
   z^2]\log z}{(x-z)^3 (z-y)^3}\nonumber\\&&\hspace{1.7cm}-\frac{2 z^2 \log x}{(x-y)
   (x-z)^3}+\frac{2 z^2 \log y}{(x-y) (y-z)^3}+\frac{x (y-3 z)+z
   (5 z-3 y)}{(x-z)^2 (y-z)^2}\Big\}.
\end{eqnarray}

The one-loop contributions to lepton EDM can be expressed as
\begin{eqnarray}
&&d_{l}^{e}=d_{l}^{\tilde{l}\chi^{0}}+d_{l}^{\tilde{\nu}\chi^{\pm}},
\nonumber\\&&d_{l}^{\tilde{l}\chi^{0}} \simeq d_l^{(a)}(\widetilde{e}_L,\widetilde{e}_R,\widetilde{B})+d_l^{(a)}(\widetilde{e}_L,\widetilde{e}_R,\lambda_{\widetilde{B'}})
+d_l^{(b)}(\widetilde{e}_R,\widetilde{H}^0,\widetilde{B})+d_l^{(b)}(\widetilde{e}_R,\widetilde{H}^0,\lambda_{\widetilde{B'}})
\nonumber\\&&\hspace{1.3cm}+d_l^{(c)}(\widetilde{e}_L,\widetilde{H}^0,\widetilde{B})+d_l^{(c)}(\widetilde{e}_L,\widetilde{H}^0,\lambda_{\widetilde{B'}})
+d_l^{(d)}(\widetilde{e}_L,\widetilde{H}^0,\widetilde{W}^0)\nonumber\\&&\hspace{1.3cm}+d_l^{(e)}(\widetilde{e}_L,\widetilde{e}_R,\lambda_{\widetilde{B'}},\widetilde{B})
+d_l^{(e)}(\widetilde{e}_L,\widetilde{e}_R,\widetilde{B},\lambda_{\widetilde{B'}}),
\nonumber\\&&d_{l}^{\tilde{\nu}\chi^{\pm}} \simeq d_l^{(f)}(\widetilde{W}^\pm,\widetilde{H}^\pm,\widetilde{\nu}_L^I)+d_l^{(f)}(\widetilde{W}^\pm,\widetilde{H}^\pm,\widetilde{\nu}^R_L).
\end{eqnarray}

\subsection{The Neutron EDM}
The neutron EDM can also be obtained from the effective Lagrangian Eq.(\ref{eq1}). The neutron is a composite particle composed of quarks and gluons. Its EDM arises not only from the quark EDM but also receives important contributions from the quark CEDM. This contribution is described by the operator $\bar{q} T^a \sigma_{\mu\nu} \gamma_5 q G_a^{\mu\nu}$, where $G_a^{\mu\nu}$ is the gluon field strength tensor.

\begin{figure}[h]
\setlength{\unitlength}{5mm}
\centering
\includegraphics[width=5in]{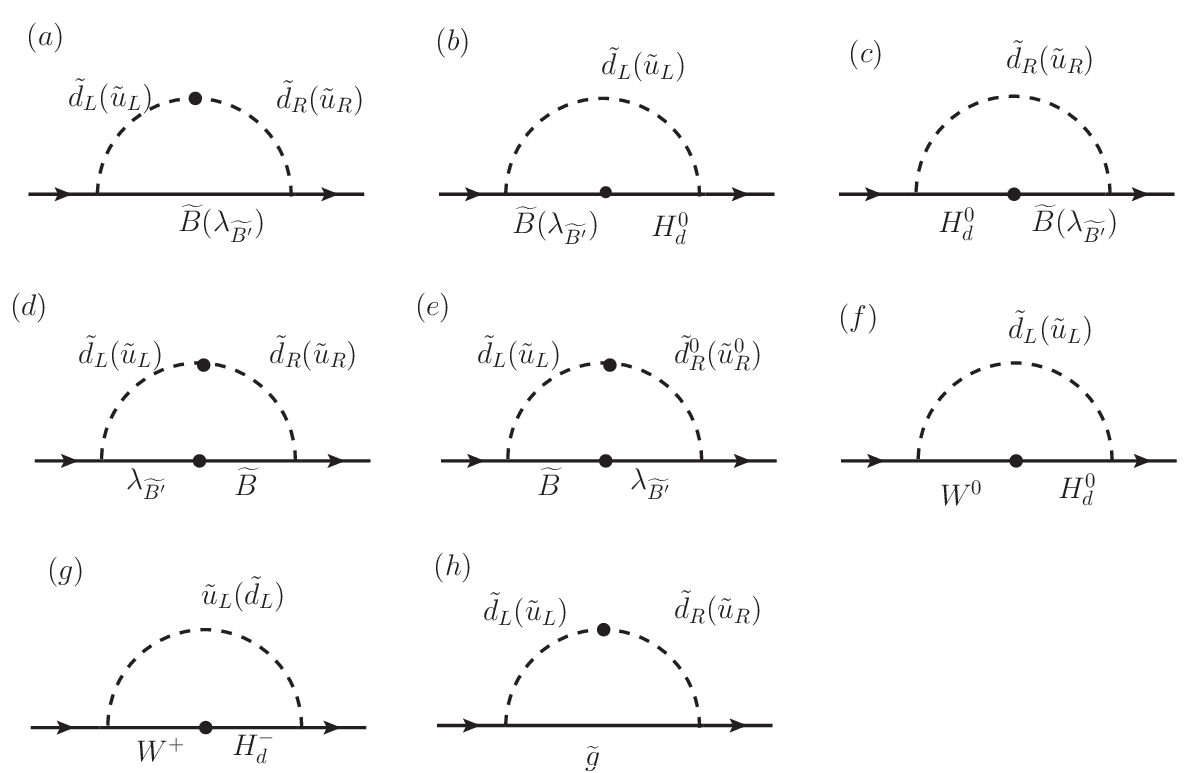}
\caption{Quark Feynman diagrams contributing to neutron EDM and CEDM generation under the MIA. External photons are attached to charged internal lines through all allowed insertion configurations, whereas external gluons couple to the colored internal lines of the loop structures.}\label{T21}
\end{figure}

The one loop diagrams for generating neutron EDM and CEDM based on MIA are shown in Fig.\ref{T21}.
Fig.~1(e) and Fig.~2(d), (e) differ from the other one-loop diagrams. These diagrams involve two distinct neutral gauginos connected through the soft mixing mass parameter $M_{BB'}$. In the N-B-LSSM, $M_{BB'}$ is a special parameter, which represents a remarkable feature of the N-B-LSSM compared with the MSSM. The existence of $M_{BB'}$ gives rise to additional contributions related to the extended neutralino sector.
Accordingly, when the relevant couplings and mass parameters carry complex phases, this class of diagrams further induces extra CP-violating contributions associated with the extended gauge sector, which is absent in the MSSM. Therefore, this class of diagrams constitutes an important ingredient of the complete EDM calculation in the N-B-LSSM.

a. The one-loop EDM contributions for quarks (u quark and d quark)
induced by the $\widetilde{B}$-$\tilde{q}_L$-$\tilde{q}_R$ and
 $\lambda_{\widetilde{B'}}$-$\tilde{q}_L$-$\tilde{q}_R$  loop diagrams are given by
\begin{eqnarray}
&&\hspace{0cm}d_q^{\gamma(a)}(\widetilde{q}_L,\widetilde{q}_R,\widetilde{B})=
-A_q\frac{eg_1^2}{27M_{SUSY}}\sqrt{x_1x_qx_{\mu_H}}e^{i\theta_1}e^{i\theta_{\mu_H}}
I_1(x_1,x_{\widetilde{q}_L},x_{\widetilde{q}_R}),
\\
&&d_q^{\gamma(a)}(\widetilde{q}_L,\widetilde{q}_R,\lambda_{\widetilde{B'}})=
B_q\frac{e(g_{YB}+g_B)}{54M_{SUSY}}
\sqrt{x_{\lambda_{\widetilde{B'}}}x_qx_{\mu_H}}e^{i\theta_1'}e^{i\theta_{\mu_H}}
I_1(x_{\lambda_{\widetilde{B'}}},x_{\widetilde{q}_L},x_{\widetilde{q}_R}),
\end{eqnarray}
with $q=u,~d$, $A_d=\tan\beta$, $A_u=4$, $B_d=(-2g_{YB}+g_B)\tan\beta$ and $B_u=-2(4g_{YB}+g_B)$.

The one-loop CEDM contributions for quarks (u quark and d quark)
induced by the $\widetilde{B}$-$\tilde{q}_L$-$\tilde{q}_R$
and $\lambda_{\widetilde{B'}}$-$\tilde{q}_L$-$\tilde{q}_R$  loop diagrams are given by
\begin{eqnarray}
&&\hspace{0cm}d_q^{g(a)}(\widetilde{q}_L,\widetilde{q}_R,\widetilde{B})=
A_q\frac{g_3g_1^2}{9M_{SUSY}}\sqrt{x_1x_qx_{\mu_H}}e^{i\theta_1}e^{i\theta_{\mu_H}}
I_1(x_1,x_{\widetilde{q}_L},x_{\widetilde{q}_R}),
\\
&&\hspace{0cm}d_q^{g(a)}(\widetilde{q}_L,\widetilde{q}_R,\lambda_{\widetilde{B'}})=
-B_q\frac{g_3(g_{YB}+g_B)}{18M_{SUSY}}
\sqrt{x_{\lambda_{\widetilde{B'}}}x_qx_{\mu_H}}e^{i\theta_1'}e^{i\theta_{\mu_H}}
 I_1(x_{\lambda_{\widetilde{B'}}},x_{\widetilde{q}_L},x_{\widetilde{q}_R}).
\end{eqnarray}

b. The one-loop EDM contributions for for quarks (u quark and d quark)
induced by the $\widetilde{B}$-$\widetilde{H}^0$-$\tilde{q}_L$ and
$\lambda_{\widetilde{B'}}$-$\widetilde{H}^0$-$\tilde{q}_L$ loop diagrams are given by
\begin{eqnarray}
&&d_q^{\gamma(b)}(\widetilde{q}_L,\widetilde{H}^0,\widetilde{B})=C_q\frac{eg_1^2}{18M_{SUSY}}
\sqrt{x_1x_qx_{\mu_H}}e^{i\theta_1}e^{i\theta_{\mu_H}}
I_2(x_1,x_{\mu_H},x_{\widetilde{q}_L});
\\
&&d_q^{\gamma(b)}(\widetilde{q}_L,\widetilde{H^0},\lambda_{\widetilde{B'}})=
-D_q\frac{e(g_{YB}+g_B)}{18M_{SUSY}}
\sqrt{x_{\lambda_{\widetilde{B'}}}x_qx_{\mu_H}}e^{i\theta_1'}e^{i\theta_{\mu_H}}
I_2(x_{\lambda_{\widetilde{B'}}},x_{\mu_H},x_{\widetilde{q}_L});
\end{eqnarray}
with $C_d=\tan\beta,~C_u=-2$, $D_d=g_{YB}\tan\beta$ and $D_u=-2g_{1}$.

The one-loop CEDM contributions for quarks (u quark and d quark) induced by the $\widetilde{B}$-$\widetilde{H}^0$-$\tilde{q}_L$ and
$\lambda_{\widetilde{B'}}$-$\widetilde{H}^0$-$\tilde{q}_L$ loop diagrams are given by
\begin{eqnarray}
&&d_q^{g(b)}(\widetilde{q}_L,\widetilde{H}^0,\widetilde{B})=-C_q\frac{g_3g_1^2}{6M_{SUSY}}
\sqrt{x_1x_qx_{\mu_H}}e^{i\theta_1}e^{i\theta_{\mu_H}}
I_2(x_1,x_{\mu_H},x_{\widetilde{q}_L}).
\\
&&\hspace{0cm}d_q^{g(b)}(\widetilde{q}_L,\widetilde{H}^0,\lambda_{\widetilde{B'}})=
D_q\frac{g_3(g_{YB}+g_B)}{6M_{SUSY}}
\sqrt{x_{\lambda_{\widetilde{B'}}}x_qx_{\mu_H}}e^{i\theta_1'}e^{i\theta_{\mu_H}}
I_2(x_{\lambda_{\widetilde{B'}}},x_{\mu_H},x_{\widetilde{q}_L}).
\end{eqnarray}

c. The one-loop EDM contributions for quarks (u quark and d quark)
induced by the $\widetilde{B}$-$\widetilde{H}^0$-$\tilde{q}_R$ and
$\lambda_{\widetilde{B'}}$-$\widetilde{H}^0$-$\tilde{q}_R$ loop diagrams are given by
\begin{eqnarray}
&&d_q^{\gamma(c)}(\widetilde{q}_R,\widetilde{H}^0,\widetilde{B})=E_q\frac{eg_1^2}{9M_{SUSY}}
\sqrt{x_1x_qx_{\mu_H}}e^{i\theta_1}e^{i\theta_{\mu_H}}
I_2(x_1,x_{\mu_H},x_{\widetilde{q}_R}),
\\
&&d_q^{\gamma(c)}(\widetilde{q}_R,\widetilde{H}^0,\lambda_{\widetilde{B'}})=
-F_q\frac{e}{18M_{SUSY}}
\sqrt{x_{\lambda_{\widetilde{B'}}}x_qx_{\mu_H}}e^{i\theta_1'}e^{i\theta_{\mu_H}}
I_2(x_{\lambda_{\widetilde{B'}}},x_{\mu_H},x_{\widetilde{q}_R}),
\end{eqnarray}
with $E_d=\tan\beta,~E_u=-4$, $F_d=g_{YB}(-2g_{YB}+g_B)\tan\beta$ and $F_u=-2g_1(4g_{YB}+g_B)$.

The one-loop CEDM contributions for quarks (u quark and d quark) induced by the $\widetilde{B}$-$\widetilde{H}^0$-$\tilde{d}_R$ and $\widetilde{B}$-$\widetilde{H}^0$-$\tilde{q}_R$ and
$\lambda_{\widetilde{B'}}$-$\widetilde{H}^0$-$\tilde{q}_R$ loop diagrams are given by
\begin{eqnarray}
&&\hspace{0cm}d_q^{g(c)}(\widetilde{q}_R,\widetilde{H}^0,\widetilde{B})=-E_q\frac{g_3g_1^2}{3M_{SUSY}}
\sqrt{x_1x_qx_{\mu_H}}e^{i\theta_1}e^{i\theta_{\mu_H}}\tan\beta
I_2(x_1,x_{\mu_H},x_{\widetilde{q}_R}).
\\
&&d_q^{g(c)}(\widetilde{q}_R,\widetilde{H}^0,\lambda_{\widetilde{B'}})=F_q
\frac{g_3}{6M_{SUSY}}
\sqrt{x_{\lambda_{\widetilde{B'}}}x_qx_{\mu_H}}e^{i\theta_1'}e^{i\theta_{\mu_H}}
I_2(x_{\lambda_{\widetilde{B'}}},x_{\mu_H},x_{\widetilde{q}_R}).
\end{eqnarray}

d. The one-loop EDM and CEDM contributions for quarks (u quark and d quark) induced by the $\lambda_{\widetilde{B'}}$-$\widetilde{B}$-$\tilde{q}_L$-$\tilde{q}_R$ and
loop diagrams are given by
\begin{eqnarray}
&&d_q^{\gamma(d)}(\widetilde{q}_L,\widetilde{q}_R,\lambda_{\widetilde{B'}},\widetilde{B})=
-E_q\frac{eg_1(g_{YB}+g_B)}{18M_{SUSY}}\sqrt{x_qx_{BB'}x_{\mu_H}}e^{i\theta_{BB'}}e^{i\theta_{\mu_H}}\tan\beta
\nonumber\\&&\hspace{2.4cm}\times[\sqrt{x_1x_{\lambda_{\widetilde{B'}}}}f(x_{BB'},x_{1},x_{\widetilde{q}_L},x_{\widetilde{q}_R})
-g(x_{BB'},x_{1},x_{\widetilde{q}_L},x_{\widetilde{q}_R})];
\\
&&d_q^{g(d)}(\widetilde{q}_L,\widetilde{q}_R,\lambda_{\widetilde{B'}},\widetilde{B})=
E_q\frac{g_3g_1(g_{YB}+g_B)}{6M_{SUSY}}\sqrt{x_qx_{BB'}x_{\mu_H}}e^{i\theta_{BB'}}e^{i\theta_{\mu_H}}
\nonumber\\&&\hspace{2.4cm}\times[\sqrt{x_1x_{\lambda_{\widetilde{B'}}}}f(x_{BB'},x_{1},x_{\widetilde{q}_L},x_{\widetilde{q}_R})
-g(x_{BB'},x_{1},x_{\widetilde{q}_L},x_{\widetilde{q}_R})].
\end{eqnarray}

e. The one-loop EDM and CEDM  contributions for quarks (u quark and d quark) induced by the $\widetilde{B}$-$\lambda_{\widetilde{B'}}$-$\tilde{q}_L$-$\tilde{q}_R$  loop diagrams are given by
\begin{eqnarray}
&&d_q^{\gamma(e)}(\widetilde{q}_L,\widetilde{q}_R,\widetilde{B},\lambda_{\widetilde{B'}})=-G_q\frac{eg_1}{36M_{SUSY}}
\sqrt{x_qx_{BB'}x_{\mu_H}}e^{i\theta_{BB'}}e^{i\theta_{\mu_H}}
\nonumber\\&&\hspace{2.4cm}\times[\sqrt{x_1x_{\lambda_{\widetilde{B'}}}}f(x_{BB'},x_{1},x_{\widetilde{q}_L},x_{\widetilde{q}_R})
-g(x_{BB'},x_{1},x_{\widetilde{q}_L},x_{\widetilde{q}_R})];
\\&&\hspace{0cm}d_q^{g(e)}(\widetilde{q}_L,\widetilde{q}_R,\widetilde{B},\lambda_{\widetilde{B'}})
=G_q\frac{g_3g_1}{12M_{SUSY}}
\sqrt{x_qx_{BB'}x_{\mu_H}}e^{i\theta_{BB'}}e^{i\theta_{\mu_H}}
\nonumber\\&&\hspace{2.4cm}\times[\sqrt{x_1x_{\lambda_{\widetilde{B'}}}}f(x_{BB'},x_{1},x_{\widetilde{q}_L},x_{\widetilde{q}_R})
-g(x_{BB'},x_{1},x_{\widetilde{q}_L},x_{\widetilde{q}_R})],
\end{eqnarray}
with $G_d=(-2g_{YB}+g_B)\tan\beta$ and $G_u=-8(4g_{YB}+g_B)$.

f. The one-loop EDM and CEDM contributions  for quarks (u quark and d quark) induced by the $\widetilde{W}^0$-$\widetilde{H}^0$-$\widetilde{q}_L$ loop diagrams are given by
\begin{eqnarray}
&&d_q^{\gamma(f)}(\widetilde{q}_L,\widetilde{H}^0,\widetilde{W}^0)=H_q\frac{eg_2^2}{6M_{SUSY}}
\sqrt{x_2x_qx_{\mu_H}}e^{i\theta_2}e^{i\theta_{\mu_H}}
I_2(x_2,x_{\mu_H},x_{\widetilde{q}_L});
\\
&&\hspace{0cm}d_q^{g(f)}(\widetilde{q}_L,\widetilde{H}^0,\widetilde{W}^0)=-H_q\frac{g_3g_2^2}{2M_{SUSY}}
\sqrt{x_2x_qx_{\mu_H}}e^{i\theta_2}e^{i\theta_{\mu_H}}
I_2(x_2,x_{\mu_H},x_{\widetilde{q}_L}).
\end{eqnarray}
with $H_d=\tan\beta$ and $H_u=2$.

g. The one-loop EDM and CEDM contributions for quarks (u quark and d quark) induced by the $\widetilde{W}^\pm$-$\widetilde{H}^\pm$-$\widetilde{q}_L$ loop diagrams are given by
\begin{eqnarray}
&&d_g^{\gamma(g)}(\widetilde{q}_L,\widetilde{W}^{\pm},\widetilde{H}^{\pm})=
I_q\frac{eg_2^2}{M_{SUSY}}\sqrt{x_2x_qx_{\mu_H}}e^{i\theta_2}e^{i\theta_{\mu_H}}
\{\frac{2}{3}I_5(x_2,x_{\mu_H},x_{\widetilde{q}_L})+\frac{1}{4}I_4(x_2,x_{\mu_H},x_{\widetilde{q}_L^0})\},
%%%%%%%%%%%%%%%%%%%%%%%%%%%%%%%%%%%%%%%%%%%%
\\
&&\hspace{0cm}d_g^{g(g)}(\widetilde{q}_L,\widetilde{W}^{\pm},\widetilde{H}^{\pm})=
I_q\frac{2g_3g_2^2}{M_{SUSY}}\sqrt{x_2x_qx_{\mu_H}}e^{i\theta_2}e^{i\theta_{\mu_H}}
I_5(x_{\widetilde{q}_L},x_2,x_{\mu_H}),
\end{eqnarray}
with $I_d=\tan\beta$, $I_u=\frac{1}{2}$.
The functions $I_4(x,y,z)$ and $I_5(x,y,z)$ are defined as
\begin{eqnarray}
&&I_4(x,y,z)=
\frac{1}{32\pi^2}\Big\{\frac{2 x [2 x^3 y + x^2 (x - 5 y) z + y (x + y) z^2] \log x}{(x - y)^3 (x - z)^3}
\nonumber\\&&\hspace{1.2cm}- \frac{2 y [2 x y^3 + y^2 (-5 x + y) z + x (x + y) z^2] \log y}{(x - y)^3 (y - z)^3}
\nonumber\\&&\hspace{1.2cm}-\frac{ (x + y) [(x^2 + 4 x y + y^2) z^2 +2 x^2 y^2]
 - x y (5 x^2 + 2 x y + 5 y^2) z + (x^2 - 6 x y + y^2) z^3}{(x - y)^2 (x - z)^2 (y - z)^2}
\nonumber\\&&\hspace{1.2cm}+ \frac{2 z^2 [x y (x + y) - 4 x y z + (x + y) z^2] \log z}{(x - z)^3 ( z-y)^3}\Big\};
\\&&I_5(x,y,z)= \frac{1} {16 \pi ^2} \Big\{\frac{x^2 \log x}{(y-x)(x-z)^2}+\frac{y^2\log y}{(x-y)(y-z)^2}-\frac{z [2 x y-z (x+y)]\log z}{(x-z)^2 (y-z)^2}
\nonumber\\&&\hspace{1.2cm}+\frac{z}{(x-z) (z-y)}\Big\}.
\end{eqnarray}

h. The one-loop EDM and CEDM contributions for quarks (u quark and d quark) induced by the $\lambda_{\widetilde{g}}$-$\widetilde{q}_L$-$\widetilde{q}_R$  loop diagrams are given by
\begin{eqnarray}
&&d_l^{\gamma(h)}(\widetilde{q}_L,\widetilde{q}_R,\lambda_{\widetilde{g}})=
H_q\frac{eg_3^2}{18M_{SUSY}}\sqrt{x_qx_{\mu_H}x_{\lambda_{\widetilde{g}}}}e^{i\theta_{\mu_H}}e^{i\theta_3}
I_1(x_{\lambda_{\widetilde{g}}},x_{\widetilde{q}_L},x_{\widetilde{q}_R});
\\
&&\hspace{0cm}d_l^{g(h)}(\widetilde{q}_L,\widetilde{q}_R,\lambda_{\widetilde{g}})=
-H_q\frac{g_3^2}{M_{SUSY}}\sqrt{x_qx_{\mu_H}x_{\lambda_{\widetilde{g}}}}e^{i\theta_{\mu_H}}e^{i\theta_g}
\nonumber\\&&\hspace{3cm}\times[\frac{1}{6}I_1(x_{\lambda_{\widetilde{g}}},x_{\widetilde{q}_L},x_{\widetilde{q}_R})
+\frac{3}{2}I_6(x_{\lambda_{\widetilde{g}}},x_{\widetilde{q}_L},x_{\widetilde{q}_R})].
%%%%%%%%%%%%%%%%%%%%%%%%%%%%%%%%%%%%%%%%%%%%
\end{eqnarray}

The function $I_6(x,y,z)$ is defined as
\begin{eqnarray}
&&I_6(x,y,z)= \frac{1}{32\pi^2}\Big\{-\frac{2 [x^2 (x^2-xy+y^2)-x y z (x+y)+y^2 z^2]\log x}{(x-y)^3(x-z)^3}
\nonumber\\&&\hspace{1.8cm}-\frac{2 [x^2 y-x z (y+z)+yz^2]\log z}{(x-z)^3 (y-z)^2}+\frac{2 y[x^2-2 x y+y (2 y-z)]\log y}{(x-y)^3(y-z)^2}
\nonumber\\&&\hspace{1.8cm}-\frac{2 x^3-x^2 (5 y+z)+x (y^2+4 yz+z^2)+y z (y-3 z)}{(x-y)^2 (x-z)^2 (y-z)}\Big\}.
\end{eqnarray}

We define the parameters that can have CP-violating phases as follows
\begin{eqnarray}
&&m_{\tilde{B}}=M_1=|M_1|e^{i*\theta_1},
~~~m_{\tilde{W}^0}=m_{\tilde{W}^{\pm}}=M_2=|M_2|e^{i*\theta_2},
\\ \nonumber&&\mu_H=\frac{\lambda v_S}{\sqrt{2}}=M_{\mu_H}=|M_{\mu_H}|e^{i*\theta_{\mu_H}},
~~~m_{\lambda_{\widetilde{B'}}}=M_{BL}=|M_{\lambda_{\widetilde{B'}}}|e^{i*\theta_1'}, \nonumber\\&&
M_{BB'}=|M_{BB'}|e^{i*\theta_{BB'}},
~~~m_{g}=|M_{3}|e^{i*\theta_{3}},
~~~x_i=\frac{|M_i|^2}{M_{SUSY}^2}.
\end{eqnarray}
$\theta_1$, $\theta_2$, $\theta_{\mu_H}$, $\theta_{1'}$ and
 $\theta_{BB'}$ are the CP-violating phases.

For the neutron EDM, the quark EDM and CEDM both give contributions. We use the following formula \cite{dn1}
\begin{eqnarray}
d_n=g^u_Td_u+g^d_Td_d+g^s_Td_s+1.1e(0.5\tilde{d}_u+\tilde{d}_d)~.
\label{dnn}
\end{eqnarray}
 Using lattice QCD \cite{dn2} at a renormalization scale of 2 GeV,
 the values of tensor charges have been obtained as $g^u_T=-0.233(28),~g^d_T=0.774(66)$ and $g^s_T=0.009(8)$.
 To improve the reliability of the results, the authors calculated it
 again at a renormalization scale of 2 GeV using QCD sum rules\cite{dn3}.
  The results obtained from two approaches (lattice QCD and QCD sum rules) have some differences\cite{dn4}, which may affect
the coefficients in Eq.(\ref{dnn}) weakly.
 In the end, it can not influence the results qualitatively.
 For neutron EDM, we can safely neglect the corrections
 from the CEDMs of heavy quarks. Even the contribution from
 s quark can be neglected safely, because $g_T^s\ll g_T^d$.

\section{Numerical analysis}
In our previous work \cite{WXJPG},
under the framework of BLMSSM we have compare MIA and mass eigenstate numerically
in detail for muon anomalous magnetic moment.
With different varying parameters in large regions,
the relative error $|\frac{a_\mu^{MIA}-a_\mu^{ME}}{a_\mu^{MIA}}|\leq 10\%$.
Within the common range of values, the relative error is not larger than $5\%$.
In this work, the relative error between the MIA results and mass eigenstate results
are in the region $2.5\%\sim 5.0\%$. In the whole, the MIA is reliable.

When performing the calculations,
the following experimental limitations are taken into account.

1. The experimental constraint from the lightest CP-even Higgs
mass $125.20\pm0.11$ {\rm GeV} is considered\cite{A8,25s33,26s34}.

2. The $Z^\prime$ boson mass should be larger than 5.1 TeV.
The ratio between $M_{Z^\prime}$ and its gauge $M_{Z^\prime}/g_B$ is not smaller than $6 ~{\rm TeV}$\cite{27s42}.

3. The new angle $\beta_\eta$ is constrained by LHC as $\tan\beta_\eta<1.5$ \cite{tanbetan}.

4. The limitations for the particle masses are the following\cite{28g43,29g44,30g45,31g46,32g47,33g48}. The neutralino mass is limited to more than 116 GeV, the chargino mass is limited to more than 900 GeV
and the slepton masses are near TeV order. The squark mass is not smaller than 1500 GeV.
The gluino mass is larger than 2000 GeV.

5. The experimental results of $\bar{B}\to X_s\gamma$ constrain strictly on the value of $\tan\beta$. The Yukawa coupling of the bottom quark reads $Y_b=\sqrt{2(1+\tan^2\beta)}\,\frac{m_b}{v}.$
With the parameter values $m_b\simeq4.18\ \mathrm{GeV}$ and $v\simeq246\ \mathrm{GeV}$,
we need $\tan\beta \lesssim 40$ to satisfy the condition $Y_b<1$.

In this work to obtain the numerical results, the relevant parameters are shown here.
$v_{S}=2~{\rm TeV}$,
$\lambda=0.25$,
$M_{SUSY}=1500~{\rm GeV}$,~$m_{\tilde{W}^0}=900~{\rm GeV}$,
$m_{\tilde{u}_L}=1800~{\rm GeV}$,
~$m_{\tilde{u}_R}=2000~{\rm GeV}$.
\subsection{Numerical analysis for lepton EDM}
Leptons are elementary particles. Their EDMs directly characterize CP violation effects in fundamental interactions and strongly depend on the lepton masses. The $\tau$-lepton, the heaviest lepton, is most sensitive to CP violation. The electron, the lightest lepton, has an extremely weak EDM signal that requires complex detection. The $\mu$-lepton resides in the intermediate response range. By adopting the MIA, the EDM can be expressed as an analytical combination of various mass insertion parameters and the imaginary parts of CP-violating phases. Thus, it shows clear parametric responses to gauge coupling constants $g_{\text{YB}}$, $g_{\text{B}}$, the Higgs vacuum structure $\tan\beta$, and independent CP phases $\theta_{\mu_H}$, $\theta_{1}$.

\subsubsection{The tau EDM}
\begin{figure}[ht]
\setlength{\unitlength}{5mm}
\centering
\includegraphics[width=3.1in]{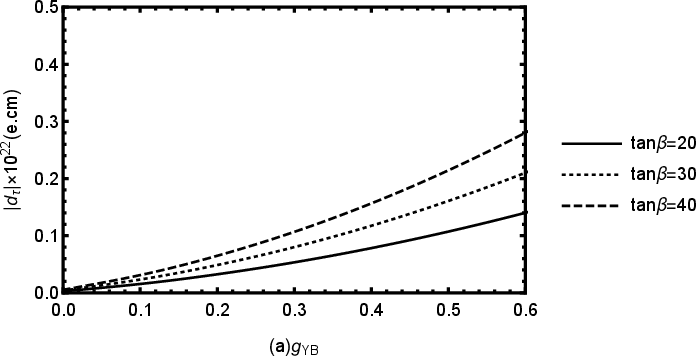}
\vspace{0.2cm}
\setlength{\unitlength}{5mm}
\centering
\includegraphics[width=3.1in]{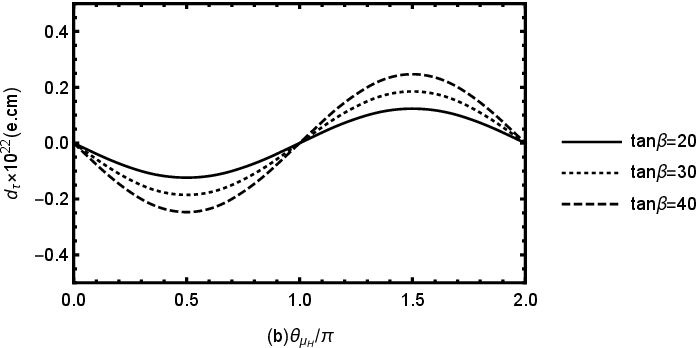}
\caption{The effects of $g_{YB}$, $\theta_{\mu_H}$ and $\tan\beta$ on $d_{\tau}$.}{\label{T3}}
\end{figure}

Fig.\ref{T3} (a) shows the variation of $|d_\tau|$ with $g_{YB}$ for different values of $\tan\beta$. The phase $\theta_{\mu_H}$=$\frac{\pi}{3}$ and other phases $\theta_{1}=\theta_{2}=\theta_{1'}=\theta_{BB'}=0$. In the range of $0 \le g_{YB} \le 0.60$, $|d_\tau|$ exhibits an approximately linear increase with $g_{YB}$, indicating that this gauge interaction has an enhancing effect on the EDM of the $\tau$ lepton. At the same value of $g_{YB}$, increasing $\tan\beta$ leads to an overall increase in $|d_\tau|$. Fig.\ref{T3} (b) presents the  variation of $d_\tau$ with $\theta_{\mu_H}$ for different $\tan\beta$. Other phases $\theta_{1}=\theta_{2}=\theta_{1'}=\theta_{BB'}=0$. $d_\tau$ undergoes a standard sinusoidal oscillation with a period of $2\pi$ as $\theta_{\mu_H}$ varies, with zeros located at $\theta_{\mu_H}=0, \pi, 2\pi$. As $\tan\beta$ increases, the oscillation amplitude increases significantly while the zero-crossing positions remain unchanged. This indicates that $\tan\beta$ does not change the phase structure, but only affects the weight of the phase term in the total amplitude. $\theta_{\mu_H}$ provides a new CP-violating source independent of the CKM phase in the SM. The obtained values of $d_\tau$ are consistent with the current experimental constraints, being approximately three orders of magnitude lower than the experimental upper limit.

\subsubsection{The muon EDM}
\begin{figure}[ht]
\setlength{\unitlength}{5mm}
\centering
\includegraphics[width=3.1in]{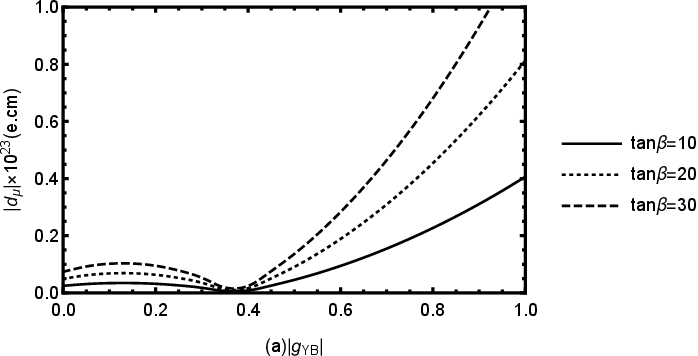}
\vspace{0.2cm}
\setlength{\unitlength}{5mm}
\centering
\includegraphics[width=3.1in]{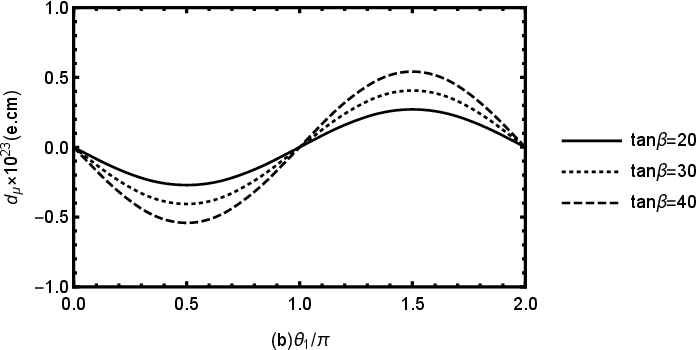}
\caption{The effects of $g_{YB}$, $\theta_{1}$ and $\tan\beta$ on $d_{\mu}$.}{\label{T4}}
\end{figure}

Fig.\ref{T4} (a) illustrates the variation of $|d_\mu|$ with $|g_{YB}|$ for different values of $\tan\beta$. Phase $\theta_{\mu_H}$=$\frac{\pi}{3}$ and other phases $\theta_{1}=\theta_{2}=\theta_{\mu_H}=\theta_{1'}=\theta_{BB'}=0$. In the range of $0 \leq |g_{YB}| \leq 1.0$, $|d_\mu|$ first increases slightly and then decreases to zero as $|g_{YB}|$ increases. This indicates that there exists a mutual cancellation effect between the contributions of different loop diagrams within this parameter range. When $|g_{YB}| \in [0.4, 1.0]$, $|d_\mu|$ resumes a monotonically increasing trend, and the growth rate accelerates with the increase of $\tan\beta$. Fig.\ref{T4} (b) depicts the variation behavior of $d_\mu$ with the phase $\theta_1$ for $\tan\beta = 20, 30, 40$. Other phase $\theta_{2}=\theta_{\mu_H}=\theta_{1'}=\theta_{BB'}=0$. Similar to the case of the $\tau$ lepton, $d_\mu$ exhibits a typical sinusoidal oscillation with a period of $2\pi$ as $\theta_1$ varies, where the oscillation zeros correspond to $\theta_1 = 0, \pi, 2\pi$. This reflects the periodic dependence of this CP-violating phase in amplitude interference. With the increase of $\tan\beta$, the oscillation amplitude is significantly enhanced. This demonstrates that $\tan\beta$ does not alter the phase structure itself, but rather enhances the observable strength of the CP-violating effect by boosting the relevant Yukawa couplings. As a CP-violating source independent of the CKM phase in the SM, the considerable signal induced by $\theta_1$ in the lepton EDM provides a promising avenue for probing new physics CP phases in higher-precision experiments.

\subsubsection{The electron EDM}

The experimental upper limit on the electron EDM ($d_e$) is very strict. To ensure that the predicted $d_e$ values from the N-B-LSSM  within the framework of the MIA satisfy this constraint, three commonly used suppression methods are adopted as follows: 1. Adopting small CP-violating phases; 2. Increasing the masses of relevant new particles to suppress loop amplitudes; 3. Exploiting internal cancellations among contributions from distinct phases, thereby reducing the imaginary part contributions without violating the phase structure.

\paragraph{small CP-violating phase}

\begin{figure}[ht]
\setlength{\unitlength}{5mm}
\centering
\includegraphics[width=3in]{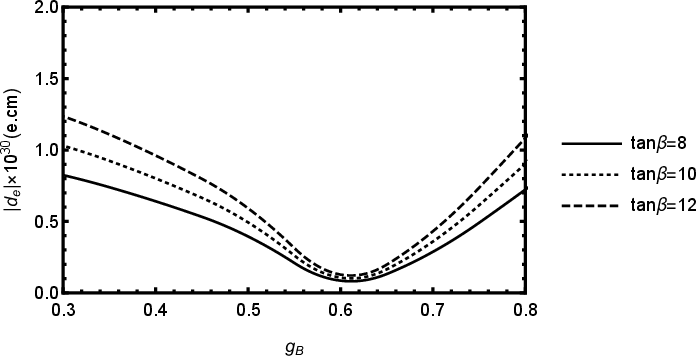}
\caption{The effects of $g_{B}$ and $\tan\beta$ on $d_{e}$.}{\label{T5}}
\end{figure}

Fig.\ref{T5} depicts the behavior of the electron EDM $|d_e|$ under the small CP-violating phase. In this case $\theta_{1'}= \frac{\pi}{1000}$, other phases $\theta_{1}=\theta_{2}=\theta_{\mu_H}=\theta_{BB'}=0$. In the figure, as $g_B$ increases from 0 to approximately 0.6, $|d_e|$ decreases monotonically and reaches a minimum at $g_B \approx 0.6$.
This indicates that the increase in $g_B$
changes the relative weights of the various loop contributions, thereby suppressing the $|d_e|$ within this region. Beyond $g_B \approx 0.6$, $|d_e|$ increases significantly, demonstrating that $g_B$ has a crucial influence on $|d_e|$ in the small phase angle limit. Different values of $\tan\beta$ (8, 10, 12) only scale the overall amplitude of $|d_e|$ at each $g_B$, without changing the overall trend.

\paragraph{large-mass approach}

\begin{figure}[ht]
\setlength{\unitlength}{5mm}
\centering
\includegraphics[width=3in]{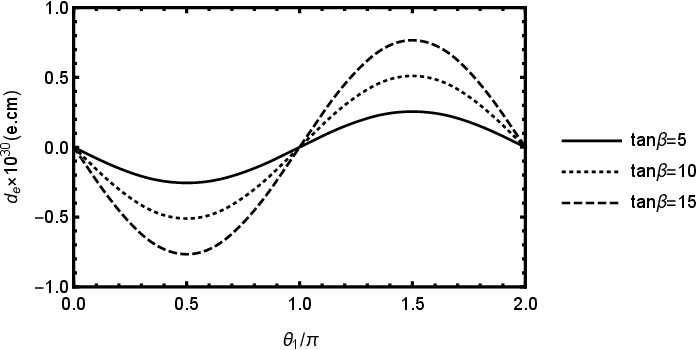}
\caption{The effects of $\theta_1$ and $\tan\beta$ on $d_{e}$.}{\label{T6}}
\end{figure}

In Fig.\ref{T6}, we calculate with the large-mass approach ($m_{\widetilde{e_L}}=13000\ \text{GeV}$, $m_{\widetilde{e_R}}=14000\ \text{GeV}$) for $\tan\beta=5$, 10, and 15. The phase $\theta_{2}=\theta_{\mu_H}=\theta_{1'}=\theta_{BB'}=0$. The electron EDM $d_e$ still exhibits sinusoidal oscillations with a $2\pi$ period as a function of $\theta_1$, consistent with the patterns observed for the $\tau$-lepton and $\mu$-lepton. However, the oscillation amplitude is extremely small. Heavy slepton only reduce the magnitude of $d_e$ without destroying the phase effect. Slepton masses at the 10 TeV order combined with a normal phase can satisfy the experimental constraints on $d_e$. Nevertheless, this violates the principle of naturalness.

\paragraph{the phase cancellation mechanism}

\begin{figure}[ht]
\setlength{\unitlength}{5mm}
\centering
\includegraphics[width=3in]{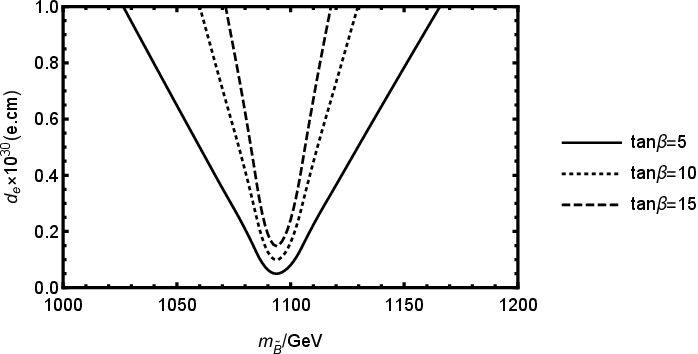}
\caption{The effects of $m_{\tilde{B}}$ and $\tan\beta$ on $d_{e}$.}{\label{T7}}
\end{figure}

In Fig.\ref{T7}, we introduce the phase cancellation mechanism and scan the mass parameter $m_{\tilde{B}}$. In this case, the phase $\theta_1=\frac{\pi}{20}$, $\theta_{1'}=\frac{\pi}{20}$, $\theta_{2}=\theta_{\mu_H}=\theta_{BB'}=0$. $d_e$ shows a clear V-shaped distribution, with a near-zero optimal cancellation point appearing around $m_{\tilde{B}}\simeq1100\ \text{GeV}$. As $\tan\beta$ decreases, the V-shaped structure generally becomes deeper and wider. This means $\tan\beta$ raises the weight of phase-related terms in the total amplitude. It can be seen that this possible method has CP-violating phases of normal size, and the particle masses are in the TeV range. The results obtained are also more likely to meet the experimental limits.
\subsection{Numerical analysis for neutron EDM}

Leptons are fundamental particles, whereas neutrons are composite particles.
Compared with the lepton EDM, the neutron EDM originates
from CP-violating interactions at the quark and gluon level
and receives contributions from quark EDM and CEDM.
Consequently, it exhibits more complicated dependencies on the model parameters.

\begin{figure}[ht]
\setlength{\unitlength}{5mm}
\centering
\includegraphics[width=3.1in]{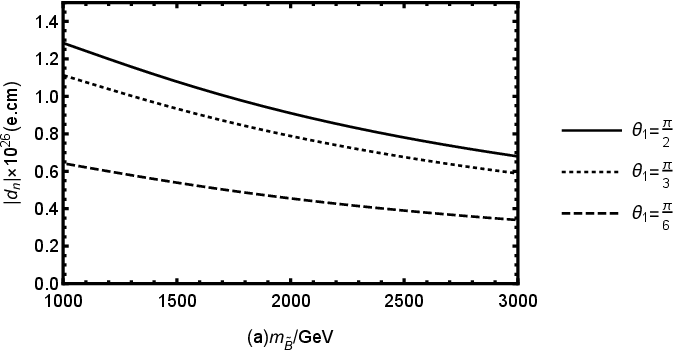}
\vspace{0.2cm}
\setlength{\unitlength}{5mm}
\centering
\includegraphics[width=3.1in]{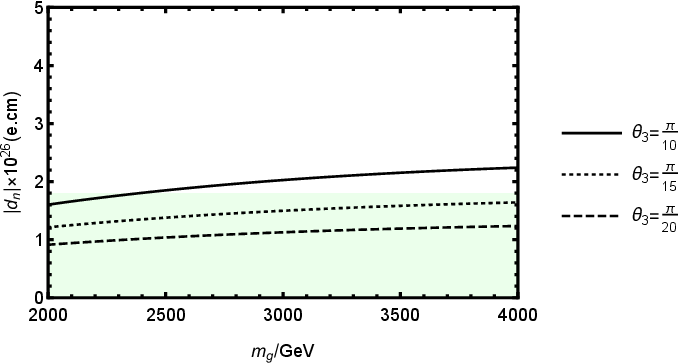}
\caption{The effects of $m_{B}$ and $\theta_{1}$ on $d_{n}$, $m_{g}$ and $\theta_{3}$ on $d_{n}$.}{\label{T8}}
\end{figure}

In Fig.~\ref{T8}(a), we study the dependence of the neutron EDM
$|d_n|$ on the $m_{\tilde{B}}$. The CP phases are fixed as $\theta_{2}=\theta_{3}=\theta_{\mu_H}=\theta_{1'}=\theta_{BB'}=0$,
with only the CP-violating phase $\theta_1$ retained. It is found that
$|d_n|$ decreases monotonically with increasing $m_{\tilde{B}}$ for different values
of $\theta_1$. The phase $\theta_1$ mainly affects the overall magnitude of
the EDM, while the dependence on $m_{\tilde{B}}$ remains unchanged. The decrease of
$|d_n|$ with increasing $m_{\tilde{B}}$ can be attributed to the mass suppression of
the loop contributions, which indicates the decoupling behavior.
Fig.~\ref{T8}(b) shows the dependence of $|d_n|$ on the gluino mass
$m_g$ for $\theta_3=\frac{\pi}{10},\frac{\pi}{15},\frac{\pi}{20}$,
where the other CP phases are set to $\theta_{1}=\theta_{2}=\theta_{\mu_H}=\theta_{1'}=\theta_{BB'}=0$.
The neutron EDM increases with $m_g$, and a larger $\theta_3$ leads to a
larger value of $|d_n|$. The increasing trend is gradually
declining. This indicates that the gluino-related contribution is sensitive to the CP-violating phase $\theta_3$. The shaded region denotes the experimental allowed range of the neutron EDM, which provides constraints on the relevant parameter space of the N-B-LSSM.

\begin{figure}[ht]
\setlength{\unitlength}{5mm}
\centering
\includegraphics[width=3.1in]{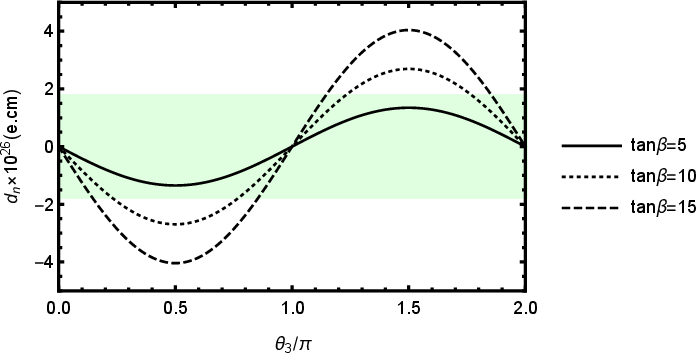}
\vspace{0.2cm}
\setlength{\unitlength}{5mm}
\centering
\includegraphics[width=3.1in]{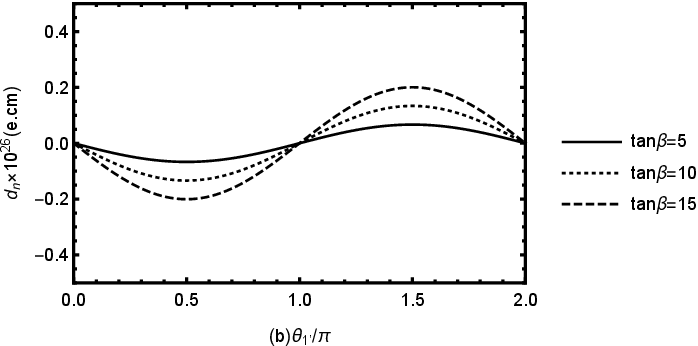}
\caption{The effects of $\theta_3$ and $\tan\beta$ on $d_{n}$, $\theta_{1'}$ and $\tan\beta$ on $d_{n}$.}{\label{T9}}
\end{figure}

Fig.~\ref{T9}(a) shows the dependence of the neutron EDM
$d_n\times10^{26}$ on the CP-violating phase $\theta_3$
for different values of $\tan\beta$, where $\theta_{1}=\theta_{2}=\theta_{\mu_H}=\theta_{1'}=\theta_{BB'}=0$.
The neutron EDM exhibits a periodic oscillating behavior as $\theta_3$ varies from $0$ to $2\pi$. The EDM vanishes at
$\theta_3=0,\pi,2\pi$, corresponding to the CP-conserving points.
Furthermore, the oscillation amplitude increases with $\tan\beta$, indicating an enhancement of the CP-violating contribution to the neutron EDM. The shaded region denotes the experimental allowed range.
Fig.~\ref{T9}(b) presents the dependence of the neutron EDM on the
CP-violating phase $\theta_{1'}$ with $\theta_{1}=\theta_{2}=\theta_{\mu_H}=\theta_{3}=\theta_{BB'}=0$.
The EDM shows a similar periodic behavior with respect to $\theta_{1'}$.
However, the oscillation amplitude is much smaller than that induced by
$\theta_3$, indicating that the contribution associated with $\theta_{1'}$
is relatively suppressed in the neutron EDM calculation.

\begin{figure}[ht]
\setlength{\unitlength}{5mm}
\centering
\includegraphics[width=4in]{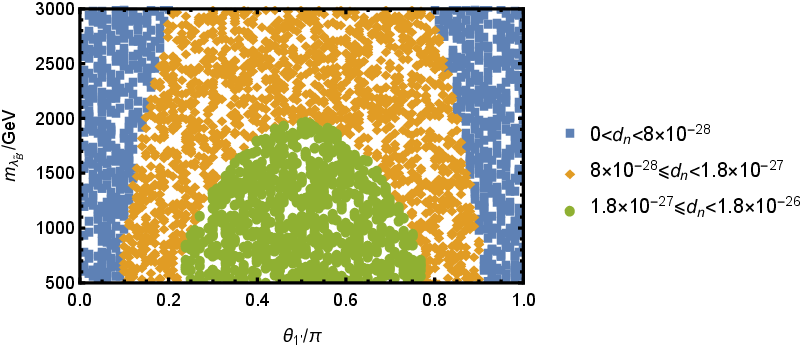}
\caption{The effects of $m_{\lambda_{\widetilde{B}'}}$ and $\theta_{1'}$ on $d_{n}$(The unit of $d_{n}$ is e.cm.), with $m_{\widetilde{d}_L}= 2100~\text{GeV}$, $m_{\widetilde{d}_R}= 1900~\text{GeV}$ and $m_g= 2500~\text{GeV}$.}{\label{T10}}
\end{figure}

Fig.~\ref{T10} presents the allowed parameter regions in the
$m_{\lambda_{B'}}-\theta_{1'}$ plane for different ranges of the neutron EDM $d_n$.
The different colors correspond to different intervals of the predicted EDM
values.
It is found that the neutron EDM is sensitive to the CP-violating phase
$\theta_{1'}$. The relatively larger values of $d_n$ are mainly located in
the regiona $\theta_{1'}\sim(0.3\pi-0.7\pi)$,
where the CP-violating contribution is enhanced. When $\theta_{1'}$ approaches the CP-conserving regions, the predicted EDM values become suppressed. In addition, the region with larger EDM values becomes narrower as $m_{\lambda_{\widetilde{B}'}}$ increases, indicating the suppression from heavy mass.

\begin{figure}[ht]
\setlength{\unitlength}{5mm}
\centering
\includegraphics[width=4in]{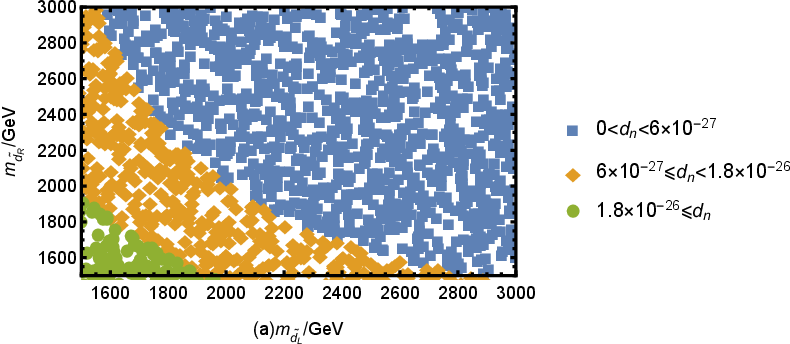}
\caption{The effects of $m_{\widetilde{d}_L}$ and $m_{\widetilde{d}_R}$ on $d_{n}$(The unit of $d_{n}$ is e.cm.), with $m_g= 2500~\text{GeV}$ and $m_{\lambda_{\widetilde{B}'}}= 3000~ \text{GeV}$.}{\label{T11}}
\end{figure}

As shown in Fig.~\ref{T11} with the CP-violating phase $\theta_1=\frac{\pi}{6}$,the other phases $\theta_{2}=\theta_{3}=\theta_{\mu_H}=\theta_{1'}=\theta_{BB'}=0$., the larger EDM values (green region) are
mainly distributed in the parameter region with relatively small
$m_{\widetilde{d}_L}$ and $m_{\widetilde{d}_R}$. When both mass parameters are small, the
corresponding supersymmetric loop contributions are less suppressed,
resulting in larger neutron EDM values. With increasing $m_{\widetilde{d}_L}$ or
$m_{\widetilde{d}_R}$, the predicted $d_n$ gradually decreases, indicating that the
larger masses suppress the relevant loop corrections. The different colored
regions in Fig.~\ref{T11} represent the predicted neutron EDM values for
different combinations of $m_{\widetilde{d}_L}$ and $m_{\widetilde{d}_R}$. By imposing the current
experimental upper bound on the neutron EDM, the parameter regions predicting
relatively large EDM values(green region) can be excluded, leading to further constraints
on the allowed ranges of $m_{\widetilde{d}_L}$ and $m_{\widetilde{d}_R}$.

\section{Conclusion}
Within the framework of the N-B-LSSM, we systematically calculate the one-loop contributions to the EDMs of leptons and quarks via the MIA, and perform a comprehensive numerical analysis. This study clearly reveals the modulating effects of kinetic mixing parameters and novel CP-violating phases on EDMs.

In the lepton sector, the coupling parameter \(g_{YB}\) can modify the relative weights of different one-loop contribution terms and induce destructive interference regions. Phase parameters of the \(\theta\)-type yield a typical behavior where EDMs oscillate periodically with a $2\pi$ cycle. In contrast, $\tan\beta$ enlarges the overall magnitude of EDMs primarily by enhancing the weights of chiral mixing, without changing the zero-point distribution structure of EDMs. Under the strict limitations of experiments, the electron EDM can be effectively suppressed through small phase values, high mass scales, or phase cancellation mechanisms. Notably, the phase cancellation mechanism allows for the existence of large CP-violating phases even when the particle mass is at the TeV scale, thereby improving the acceptability of the model.

The construction of the neutron EDM includes contributions from quark EDMs and quark CEDMs, resulting in richer parameter dependent features. $\tan\beta$ has a stable enhancing effect on the neutron EDM, whereas the newly introduced mass scale shows a suppressive trend. The strong phase $\theta_3$ usually dominates the morphological characteristics of CP interference, while other phases such as $\theta_{1'}$ constitute secondary but independent sources of CP violation.

EDM experiments exert rigorous joint constraints on the kinetic mixing coupling parameters and novel CP-violating phases in the N-B-LSSM. In particular, high-precision measurements of the electron and neutron EDMs provide a clear observational window for testing this class of $U(1)_{B-L}$ extended supersymmetric models.

With a reasonable selection of parameters, the N-B-LSSM can simultaneously satisfy the current experimental upper limits on the EDMs of leptons and neutrons, and certain parameter spaces can even approach the detection sensitivity of future experiments. The MIA demonstrates good applicability in this model, enabling a clear elucidation of the regulatory mechanisms of CP-violating phases and various coupling parameters on EDMs. This work establishes a complete analytical framework and numerical benchmark for the systematic study of EDMs in B-L extended supersymmetric models, and provides solid theoretical support for the detection of CP-violating phases in new physics via high-precision EDM experiments in the future.

\begin{acknowledgments}

This work is supported by National Natural Science Foundation of China (NNSFC)
(No.12075074), Natural Science Foundation of Hebei Province
(A2023201040, A2022201022, A2022201017, A2023201041), Natural Science Foundation of
Hebei Education Department (QN2022173), Post-graduate's Innovation
Fund Project of Hebei University (HBU2024SS042), the Project of the China
Scholarship Council (CSC) No. 202408130113. X. Dong acknowledges support from Funda\c{c}\~{a}o para a Ci\^{e}ncia e a Tecnologia (FCT, Portugal) through the projects CFTP FCT Unit UIDB/00777/2020 and UIDP/00777/2020.

\end{acknowledgments}

\end{document}